\documentclass
[preprint,twocolumn,prl,10pt,nofootinbib,tightenlines,bibnotes]{revtex4}%
\usepackage{amsfonts}
\usepackage{amsmath}
\usepackage{amssymb}
\usepackage{graphicx}%
\begin{document}
\preprint{UATP/08-02}
\title{Irreversibility, Molecular Chaos, and A Simple Proof of the Second Law }
\author{P. D. Gujrati}
\email{pdg@arjun.physics.uakron.edu}
\affiliation{The Department of Physics, The University of Akron, Akron, OH 44325}
\date{\today}

\begin{abstract}
The irreversibility in a statistical system is traced to its probabilistic
evolution, and the molecular chaos assumption is not its unique consequence as
is commonly believed. Under the assumption that the rate of change
$\overset{\cdot}{p_{i}}(t)$ of the $i$th microstate probability $p_{i}(t)$
vanishes only as $t\rightarrow\infty,$ we prove that the entropy of a system
at constant energy cannot decrease with time.

\end{abstract}
\maketitle

\begin{figure}
[ptb]
\begin{center}
\includegraphics[
trim=1.066648in 0.000000in 0.000000in 0.000000in,
height=1.9995in,
width=2.8643in
]%
{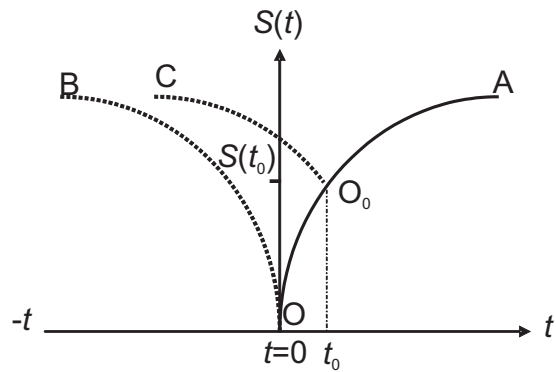}%
\caption{Schematic behavior of $S(t)$ as a function of time $t$. Starting at O
($t=0$), OA and OB show the symmetric growth of $S(t)$ in future and under
time reversal at $t=0$. If we reverse time later at $t=t_{0}+t^{\prime}$ by
setting $t^{\prime}\rightarrow-t^{\prime},$ then O$_{0}$C shows the growth of
the entropy above its value $S(t_{0})$ at $t=t_{0}$; the entropy does not
retrace O$_{0}$O, as would be required by time-reversal invariance. }%
\label{fig_St_0.ps}%
\end{center}
\end{figure}

One of the outstanding problems in theoretical physics is the demonstration of
thermodynamic irreversibility of Clausius's second law of thermodynamics
\cite{Clausius}. Its resolution is vital as the second law, whether treated as
an axiom or as a law, determines the way Nature evolves. What makes the second
law so unique is that its irreversibility is in stark contrast with
reversibility obeyed by classical or quantum mechanics governing all processes
in Nature; the only known exception is some Kaon decay over a short period of
time. As Loschmidt \cite{Loschmidt} argued, thermodynamic irreversibility
contradicts the reversibility principle; thus, we may have to abandon one of
them. Nature may be perfect, but our description of any \emph{macroscopic}
\emph{part} of it requires a \emph{probabilistic approach }due to external
noise; see later. It was first pointed out by Kr\"{o}ning \cite{Kroning}, and
later developed by Boltzmann \cite{Boltzmann0}; see \cite{ter Haar,Lebowitz}
for excellent reviews. It was the first approach in physics that established
that fundamental laws of Nature need not be strictly deterministic. Many
phenomena at the microscopic level such as the nuclear decay are also known to
require a probabilistic approach for their understanding. Thus, the
probabilistic interpretation is not just a consequence of a macroscopic nature
of the system. Nevertheless, it has to be exploited for a proper understanding
of the second law. To appreciate this, we note that the Gibbs formulation of
the entropy $S(t)$ is
\begin{equation}
S(t)=-\sum p_{i}(t)\ln p_{i}(t)\geq0,\ \ \sum p_{i}(t)\equiv
1,\label{Eq_Entropy}%
\end{equation}
where $p_{i}(t)$ is the probability of the $i$th microstate at time $t$; the
sum is over all distinct $W$ microstates. According to the second law, the
entropy of an isolated system cannot decrease with time; see OA in Fig.
\ref{fig_St_0.ps}. The probabilities change until the equilibrium is attained
\emph{asymptotically} when $S(t)\rightarrow\ln W$, its maximum possible value,
as $t\rightarrow\infty$, and all microstates have the \emph{same
}probability:
\begin{equation}
p_{i}(t)\overset{\ }{\rightarrow}1/W\text{ \ and \ }\overset{\cdot}{p_{i}%
}(t)\rightarrow0\text{ \ only when \ }t\rightarrow\infty,\text{\ }%
\label{Eq_Limiting_Value_p_i}%
\end{equation}
where the dot denotes the derivative with respect to time.%

It is not just the microstates themselves, but their probabilities of
occurrence that determine the entropy. As we demonstrate here, not
appreciating this fact has given rise to the irreversibility paradox of
Loschmidt. There have been several attempts to explain irreversibility such as
the Boltzmann $H$-theorem, coarse-graining, the Brussels School of Prigogine,
to name a few; see \cite{Coveney} for an excellent summary of these
approaches. In particular, Boltzmann argued that time-reversibility remains
intact, except that the time required to observe it is so long that it will
never occur in one's lifetime. This is the customary mechanical approach taken
in statistical mechanics for an isolated system, where statistical methods
(such as the use of ensembles, molecular chaos, etc.) are used only as a
convenience for treating a macroscopic system, while deeply believing in its
time-reversible evolution as a mechanical system. This is apparent from the
extensive use of the Liouville theorem as a framework in developing
statistical mechanics as a tool \cite{Landau}. Our \emph{stochastic approach}
will deviate from this customary mechanical approach in simple but important
ways. The origin of this probabilistic behavior is the stochastic interaction
with the environment, no matter how weak \cite{Landau,GujratiPoincare}, whose
immediate consequence is to make interparticle collisions also stochastic. We
only consider very weak environmental interaction; otherwise, separating the
system from the environment will not be useful \cite{Landau,GujratiPoincare}.
The introduction of \emph{molecular chaos }(MC) by Burbury \cite{Burbury} and
later adopted by Boltzmann \cite{BoltzmannLectures} to explain
irreversibility, however, elevates MC as the source for irreversibility.
Unfortunately, the use of MC helps to conceal, rather than reveal, the true
origin of irreversibility, as we will argue below. We will establish that (i)
irreversibility is caused by stochasticity in the system which arise from
interactions with the environment, (ii) MC is one of many possible
manifestations of it, and not the only one, and (iii) the entropy cannot
decrease if we assume (\ref{Eq_Limiting_Value_p_i}).

\begin{figure}
[ptb]
\begin{center}
\includegraphics[
trim=0.637941in 4.816343in 4.019113in 2.892360in,
height=2.6688in,
width=3.1695in
]%
{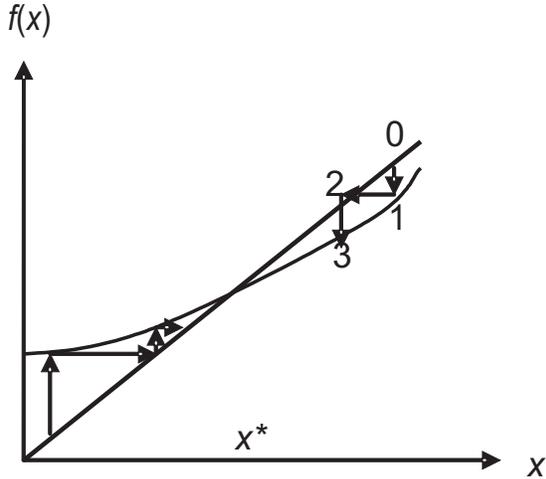}%
\caption{The function $f(x),$ $\varphi\neq0,$ shown schematically as the thin
curve, while thick straight line at 45$^{\text{o}}$ angle represents $x$. To
find the fix point $x^{\ast}$, one follows the arrows $0\rightarrow
1\rightarrow2\rightarrow3,\cdots,$ starting at $0.$ The time evolution is due
to interactions with the exterior ($\varphi\neq0$). When there is no external
interaction ($\varphi=0$), the state never changes with time, as expected. It
is clear that it will take infinte number of iterations of the recursion
relation $x_{j+1}=f(x_{j})$ to find the fix point. Under time-reversal, one
follows the arrows in reverse $\cdots,3\rightarrow2\rightarrow1\rightarrow0$
to retrieve the initial point $0$. }%
\label{fig_kac.ps}%
\end{center}
\end{figure}

The significance and relevance of our stochastic approach to study system's
evolution is convincingly demonstrated by the very simple but highly
celebrated Kac ring model \cite{Kac,Thompson}, which contains $N$\ balls of
two colors A and B, localized on sites of a ring; there are no empty sites. A
microstate $i$ represents an ordered sequence of the colors of the balls. The
time evolution occurs by balls moving in unison one step clockwise during each
time interval $\Delta$. To include the environmental effects, we consider
$F$\ flippers with\emph{ fixed} positions on the links between neighboring
sites that flip the colors of balls (A$\Leftrightarrow$B) without any bias as
they pass through the flippers. The movement of each ball is
\emph{deterministic}, even when it passes through a flipper. In one time step,
the microstate $i$ evolves into another unique microstate $i^{\prime}.$ We
denote this time evolution by a \emph{one-to-one }mapping $i\rightarrow
i^{\prime}$. It can be inverted to give the backward evolution $i^{\prime
}\leftarrow i$ under time reversal ($t\rightarrow-t$) by balls moving
counter-clockwise; there is no effect on flippers' ability to flip colors
under reversal as the balls pass through them. Thus, time reversal will
generate uniquely any past state of the system so there is \emph{time-reversal
invariance}. The number of distinct microstates of balls is $W=2^{N}$. For
very weak external interactions, we need the flipper density $\varphi\equiv
F/N\ll1$. Let $A_{j}$ and $B_{j\text{ }}$denote the number of A and B balls
and $a_{j}$ and $b_{j\text{ }}$the number of A and B balls with a flipper
ahead of them at time $t_{j}\equiv j\Delta;$ the index $j$ should not be
confused with $i$ for a microstate$.$ Obviously, $A_{j}+B_{j\text{ }}\equiv
N$, and $a_{j}+b_{j\text{ }}\equiv F$, so that we can treat only one species
(we choose B) as independent. It is easy to establish the following recursion
relation (RR) $B_{j+1}\equiv B_{j}+a_{j}-b_{j}.$ Introducing the densities
$\widetilde{P}_{j}=B_{j}/N$ and $\widetilde{p}_{j+1}=b_{j}/F,$ we can rewrite
the RR for $\widetilde{P}_{j}$:
\begin{equation}
\widetilde{P}_{j+1}\equiv\widetilde{P}_{j}+\varphi(1-2\widetilde{p}_{j}).
\label{Deterministic_Kac_Model}%
\end{equation}
This RR cannot be solved in a closed form, since it involves the quantity
$\widetilde{p}_{j}$\ which is determined by the initial microstate and $j$. To
proceed further, we need to supplemented it by some known $\widetilde{p}_{j}$.
The most direct way is to express $\widetilde{p}_{j}$\ as a function of
$\widetilde{P}_{j}$.\ We will call such an assumed relationship the
\emph{reduction assumption}. One such choice, following Burbury and Boltzmann,
is that of MC, which for the current model leads to $\widetilde{p}_{j}%
\equiv\widetilde{P}_{j}$ \cite{Kac,Thompson}:\ the density of any color is
independent of whether a flipper is ahead or not. However, many other forms
can be utilized. We will choose the following form, merely for simplicity,
\begin{equation}
1-2\widetilde{p}_{j}\equiv\theta(\varphi,\widetilde{P}_{j})\left[
1-2\widetilde{P}_{j}\right]  ; \label{Eq_Reduction_Assumption}%
\end{equation}
here, $\theta_{j}\equiv\theta(\varphi,\widetilde{P}_{j})$ is some
\emph{arbitrary} real positive function, although with some strong
restrictions to be detailed below. It includes MC\emph{ }$(\theta=1)$ as a
special case. It is one out of many possible choices, but is sufficient to
make our point that \emph{the molecular chaos assumption is not unique for
irreversibility to emerge}. The functional form of $\theta_{j}$ does not
depend on time explicitly; implicitly, it may depend on $j$ through
$\widetilde{P}_{j}.$ The RR now becomes $\widetilde{P}_{j+1}\equiv
f(\widetilde{P}_{j}),$ $f(x)\equiv x+\theta\varphi(1-2x)=\theta\varphi
+(1-2\theta\varphi)x\leq1$ over $0\leq x\leq1,$ where $\theta,x$ stand for
$\theta_{j}(\varphi,x),\widetilde{P}_{j}$, respectively. The fix point (FP)
$x^{\ast}$\ of the RR is given by $f(x^{\ast})=x^{\ast}$. The restriction
$f(x)\leq1$ strongly restricts $\theta.$ In addition, we require that (a)
$\theta(\varphi,x)$ has no zero over $0\leq x\leq1$, (b) the slope at
$x=x^{\ast}=1/2$ satisfies $0<f^{\prime}(x^{\ast})<1$, and (c) $f(x)$ is
monotonic$.$ Satisfying all these requirements will usually impose certain
restrictions on the allowed values of $\varphi$. For example, for
$\theta=\theta_{0}=$constant, it is easy to see that $\varphi\leq1/2\theta
_{0}$. With these requirements, the schematic form of $f(x)$ is as shown in
Fig. \ref{fig_kac.ps}; its convexity is not relevant. The RR can now be solved
recursively to obtain the limiting value $\widetilde{P}_{\text{eq}}$ of
$\widetilde{P}_{j}$ as $j\rightarrow\infty$, which represents the FP. To find
$x^{\ast}\ $of the RR $x_{j+1}=f(x_{j}),$ we proceed graphically and follow
the arrows in Fig. \ref{fig_kac.ps}, as we approach $x^{\ast}$. The starting
point $0$ may be above $x^{\ast}$, as shown in the figure, or below (not
shown); in both cases, we converge to the FP $x^{\ast}=1/2$. Thus, $x^{\ast}$
is an \emph{attractive} FP and determines the \emph{equilibrium state} in the model.%

The form of $f(\widetilde{P}_{j})$\ is determined by the dynamics of the
system; despite this, the FP, i.e., the equilibrium state is the \emph{same}
for all dynamics described by the above class of functions $f(\widetilde
{P}_{j})$. This class of $f(\widetilde{P}_{j})$\ also gives a \emph{monotonic}
decreasing (increasing) sequence of $\widetilde{P}_{j}$, which converges to
the FP $\widetilde{P}_{\text{eq}}=x^{\ast}=1/2$, if the initial state $0$ is
above (below) $x^{\ast}.$ The FP $\widetilde{P}_{\text{eq}}=1/2$ results from
our adopted unbiased dynamics. The monotonic behavior is in accordance with
our fundamental postulate (\ref{Eq_Limiting_Value_p_i})$.$ Such a behavior,
which follows from our reduction assumption, creates a contradiction with the
time reversal invariance of (\ref{Deterministic_Kac_Model}). What is even more
puzzling is that the reduction assumption has \emph{no} explicit stochastic
ingredient. Thus, the RR appears to have a completely deterministic form:\ we
can reverse the arrows in Fig. \ref{fig_kac.ps} and retrieve the entire
history $[\widetilde{P}_{j}\equiv f^{-1}(\widetilde{P}_{j+1})]$. Despite this,
we have an irreversible asymptotic ($j\rightarrow\infty$ or $t\rightarrow
\infty$) approach to equilibrium brought about by the reduction assumption.

To understand how the reduction assumption leads to irreversibility, we
proceed as follows. We change the above deterministic model to a
\emph{stochastic model} by introducing stochasticity: We make the positions of
the flippers random at each time step $\Delta$ to mimic \emph{stochastic
external interactions}. This means that there is a flipper present with a
probability $\varphi$ ahead of each ball. We also assume balls to be
\emph{uncorrelated}, and\ focus on one of the balls. Let $P_{j}$ denote the
probability that its color is B at time $t_{j};$ this probability should not
be confused with the density $\widetilde{P}_{j}$ above for the deterministic
system; there was no stochasticity earlier.\ Each flipper flips color with
probability $\theta_{j}\equiv\theta(\varphi,P_{j})$, and does not flip it with
probability $1-\theta_{j}$. Then, it is easy to see that%
\begin{align}
P_{j+1}  &  =(1-\varphi)P_{j}+\varphi(1-\theta_{j})P_{j}+\varphi\theta
_{j}(1-P_{j})\nonumber\\
&  =P_{j}+\varphi\theta_{j}(1-2P_{j}), \label{Stochastic_Kac_RR}%
\end{align}
which is identical to the RR (\ref{Deterministic_Kac_Model}) for the
deterministic model subjected to (\ref{Eq_Reduction_Assumption}) if we
identify the probability $P_{j}$ with the density $\widetilde{P}_{j}$. One can
also interpret the above RR by assuming that there is a flipper on each link
with certainty and it flips the color with probability $\varphi\theta_{j}$ and
does not flip it with probability $1-\varphi\theta_{j}$. As before, the
stochastic RR (\ref{Stochastic_Kac_RR}) converges \emph{monotonically} to
equilibrium ($P_{j}\rightarrow P_{\text{eq}}=%
\frac12
$) as \ $t_{j}\rightarrow\infty$, see Fig. \ref{fig_kac.ps}, so that there are
equal numbers of balls of the two colors. Thus, the probability of any of the
$W=2^{N}$ microstates is exactly $(1/2)^{N}$ in accordance with our
fundamental postulates (\ref{Eq_Limiting_Value_p_i}). It is clear now that the
reduction assumption (we have considered one of many possible classes of $f$)
is really a consequence of our stochastic model. A class different than
(\ref{Eq_Reduction_Assumption}) relating $\widetilde{p}_{j}$ with
$\widetilde{P}_{j}$ may result in a different equilibrium state. However, even
in this case, the dynamics will be irreversible. Thus, there is nothing unique
about the Burbury-Boltzmann MC assumption or our reduction assumption as far
as irreversibility is concerned.\ The former is really one of many conceivable
reduction forms that emerge as a consequence of an irreversible dynamics. We
should point out that (\ref{Stochastic_Kac_RR}) can be inverted for the
probabilities:$P_{j}\equiv f^{-1}(P_{j+1})$. Graphically, this requires
reversing the arrows in Fig. \ref{fig_kac.ps}, which will eventually yield the
initial probability $P_{0}.$ However, as we will see below, this inversion of
the probability RR does not correspond to time-reversal evolution of the
microstates. Thus, it says nothing about the reversibility of the second law.
For the molecular chaos case, it is easy to show
\[
P_{j}=P_{\text{eq}}+(1-2\varphi)^{j}(P_{0}-P_{\text{eq}}).
\]

Now that we have seen that MC is no longer a unique manifestation of
irreversibility, we need to look for a much deeper cause of irreversibility,
as a variety of dynamics will lead to irreversibility. Thus, a particular
dynamics itself cannot be the unique cause of irreversibility. We propose that
it is the stochastic nature of a statistical system that is responsible for
time irreversibility \cite{GujratiPoincare}. Its presence invalidates the time
reversal invariance. We now justify this proposal. Let us first consider a
deterministic evolution in which an initial microstate $i$ at time $t=0$
evolves in a unique fashion. From (\ref{Eq_Entropy}), $S(0)\equiv0$. Let us
consider times that are integral multiple $t_{j}\equiv j\Delta$ in terms of
some fixed interval $\Delta$. Then $i$ evolves into a microstate $i_{j}%
$\ ($i\rightarrow i_{j}$)\ at a future time $t_{j}$ with certainty, i.e. with
probability one, to use the language of probability. Again, using
(\ref{Eq_Entropy}), $S(t_{j})\equiv0$. The deterministic dynamics possesses
the property of \emph{time-reversal invariance}. Since the evolution
$i\rightarrow i_{j}$\ is \emph{one-to-one}, strict \emph{causality} ("from the
same antecedents follow the same consequents") is maintained and the mapping
can be \emph{inverted} at any time. Thus, the forward evolution $i\rightarrow
i_{1}\rightarrow i_{2}\rightarrow\cdots\rightarrow i_{j}$ of an initial
microstate can be uniquely inverted to give $i_{j}\leftarrow i_{j-1}\leftarrow
i_{j-2}\leftarrow\cdots\leftarrow i_{1}\leftarrow i$, and we recover the
initial microstate in this reversal. The entropy in this reversal remains
constant$=0$, which is consistent with the time-reversal invariance. The
macroscopic irreversibility observed in a macroscopic system should also not
be confused with the \emph{chaotic }behavior seen in a system with only a few
degrees of freedom, the latter being purely deterministic and, therefore,
should not be considered an example of time-irreversibility from what we said above.

For the concept of entropy to be useful requires a particular kind of
probabilistic approach in which the evolution must not be deterministic, even
though one can use densities such as $\widetilde{P}_{j}$ above for a
deterministic system as a suggestive probability; rather, it must be genuinely
$\emph{stochastic}$. As Landau observes \cite{Landau}, even an isolated system
is not truly deterministic in Nature. A real system must be confined by a real
container, which forms the exterior of the system. The container cannot be a
perfect insulator. Moreover, it itself will introduce environmental noise in
the system. Thus, there are always stochastic disturbances going on in a real
system due to the exterior, which \emph{cannot} be eliminated, though they can
be minimized. (For the Kac ring model, we must ensure $\varphi\ll1$, but we
must not have $\varphi\equiv0;$ in the latter case, the probabilities $P_{j}$
will never change and the entropy will remain constant as we have a
deterministic evolution.) For quantum systems, this requires considering the
Landau-von Neumann density matrix, rather than eigenstates \cite{Landau}. The
derivation in \cite{Landau} clearly shows the uncertainty introduced by the
presence of "outside". The latter is not being part of the system, just as the
flippers are not used in identifying the microstates and their density does
not affect the FP in the Kac model.

We do not have to consider the actual nature of the noise; all that is
required is its mere presence. One can think of $\varphi$\ in the Kac model as
the strength of stochastic noise. As long as $\varphi\neq0$ (or $\varphi
\theta\neq0$), the model will \emph{always} converge to the FP, i.e. will
equilibrate. It is only in this case that the entropy will increase as the
probabilities of various microstates change in time, as we prove below. The
actual nature of the noise will only determine the form of the dynamics, but
not the final equilibrium state, which remains oblivious to the actual noise
or the dynamics. This is what allows the statistical mechanical approach to
make predictions about the equilibrium state. Since the evolution is
stochastic, a microstate $i$ makes a "jump" to one of the $W$ microstates in
the microstate set $\left\{  I\right\}  $. The mapping $i\dashrightarrow
\left\{  I\right\}  $ is \emph{one-to-many}, with each of the possible
$i^{\prime}\in\left\{  I\right\}  $ occurring with certain probabilities.
Because of the one-to-many nature, the mapping \emph{cannot} be inverted to
study time-reversal, and strict causality is destroyed. To appreciate this
observation, let us consider the Kac ring model. Under time-reversal, the
balls move counter-clockwise, but flippers continue to flip colors with the
same probability $\varphi\theta$. Thus, $i^{\prime}$ gives rise to two
possibilities so the mapping still remains one-to-many, which causes
irreversibility:$P_{j+1}$ does not go to $P_{j}$. Thus, time reversal and
arrow reversals are not the same.\ 

Let us reverse time at $t=t_{0}>0$, where the entropy is $S(t_{0});$ see Fig.
\ref{fig_St_0.ps}. Then all possible microstate in the set $\left\{
I\right\}  $ will not uniquely jump back into $i;$\ rather each of the
possible $i^{\prime}$\ will stochastically jump to any of the $W$ microstates
in the set $\left\{  I\right\}  $, and the entropy will continue to increase
from its value $S(t_{0})$. This is shown by O$_{0}$C in Fig. \ref{fig_St_0.ps}%
. The entropy will not follow O$_{0}$O. The stochasticity destroys
time-reversal invariance of the dynamics at $t_{0}>0$. The situation is
different at $t=0,$ when the system was initially prepared in some
non-equilibrium state. If we follow its evolution in past and in future
separately, we will discover that the entropy continues to increase until
equilibrium is reached in both cases; see OA and OB in Fig. \ref{fig_St_0.ps}.
Thus, time-reversal invariance is valid with respect to $t=0$, but not when
$t>0.$

We now consider a general system, which is initially prepared so that the
initial probabilities $p_{i0}\equiv p_{i}(t=0)$ of various microstates are not
zero. From (\ref{Eq_Limiting_Value_p_i}), we see that if the initial
probabilities $p_{i0}\geq1/W$ for some microstates, then these probabilities
must decrease ($\overset{\cdot}{p_{i}}\leq0$). On the other hand, if $p_{i0}$
$<$
$1/W$ for some microstates, then these probabilities must increase
($\overset{\cdot}{p_{i}}\geq0$). Accordingly, we partition microstates into
two distinct groups $\mathcal{G}$ (containing microstates with $p_{i}\geq1/W$)
and $\mathcal{L}$ (containing microstates with $p_{i}<1/W$). We will assume
$p_{0}$ is the smallest of all probabilities in the group $\mathcal{G}$. We
treat $p_{0}$ to be the dependent variable, while all other $p_{i}$'s, $i>0,$
are treated as independent variables, so that $p_{0}\geq1/W$ is given by
$p_{0}=1-\sum_{i>0}p_{i}.~$We find that%
\[
dS/dt=\sum_{i>0}\overset{\cdot}{p_{i}}(dS/dp_{i})=\sum_{i\in\mathcal{G}%
}\overset{\cdot}{p_{i}}\ln(p_{0}/p_{i})+\sum_{i\in\mathcal{L}}\overset{\cdot
}{p_{i}}\ln(p_{0}/p_{i}).
\]
As the two factors in each of the two sums have the same sign, $dS/dt\geq0.$
This establishes that the entropy continues to increase until $\overset{\cdot
}{p_{i}}(t)\rightarrow0$ for all microstates, in which case
(\ref{Eq_Limiting_Value_p_i}) finally becomes satisfied. During time-reversal
dynamics, the derivatives $\overset{\cdot}{p_{i}}(t)$ will have changed their
signs, making $dS/dt\leq0,$ so that the second law is \emph{not} violated.

For a deterministic evolution, the entropy remains a constant of motion
\cite{GujratiPoincare}. This is consistent with the Liouville theorem
according to which the volume of the phase space is a constant of motion under
deterministic dynamics, and which also does not violate the second law. In our
opinion, the stochastic interpretation adopted here easily explains Maxwell's
idea \cite{Maxwell}that molecular motion is "perfectly irregular";\textbf{
}this irregularity must be present in order for the system to behave
irreversibly. Our approach is sufficient to show irreversibility; one does not
really need to use the oscillatory behavior of the H-curve \cite{Landau} as
proposed by Boltzmann to resolve the paradox.

\end{document}